\documentclass[aps,prl,twocolumn,showpacs,superscriptaddress,groupedaddress]{revtex4}  % for review and submission
\usepackage{graphicx}  % needed for figures
\usepackage{dcolumn}   % needed for some tables
\usepackage{bm}        % for math
\usepackage{amssymb}   % for math
\usepackage{color}
\usepackage{pdfpages}

\begin{document}

\title{Electron-photon scattering mediated by localized plasmons: A quantitative analysis by eigen-response theory}

\author{Kin Hung Fung} \affiliation{Department of Mechanical Engineering, Massachusetts Institute of Technology, Cambridge, MA 02139, USA}\affiliation{Department of Applied Physics, The Hong Kong Polytechnic University, Hong Kong}
\author{Anil Kumar} \affiliation{Department of Electrical and Computer Engineering, University of Illinois at Urbana-Champaign, Urbana, Illinois 61801, USA}
\author{Nicholas X. Fang} \email{nicfang@mit.edu}\affiliation{Department of Mechanical Engineering, Massachusetts Institute of Technology, Cambridge, MA 02139, USA}
\date{\today}

\begin{abstract}

We show that the scattering interaction between a high energy
electron and a photon can be strongly enhanced by different types of
localized plasmons in a non-trivial way. The scattering interaction is predicted by an
eigen-response theory, numerically verified by finite-difference-time-domain
simulation, and experimentally verified by cathodoluminescence spectroscopy. We find that the scattering interaction
associated with dark plasmons can be as strong as that of
bright plasmons. Such a strong interaction may offer new opportunities to improve
single-plasmon detection and high-resolution characterization techniques for
high quality plasmonic materials.

\end{abstract}

\maketitle

The strength of the near-field resonant response of plasmonic nanostructures plays an
important role in the processes of spontaneous emission
\cite{Farahani:2005,Taminiau:2008,Choy:2011} and
stimulated emission \cite{Berini:2012,Bergman:2003,Hill:2009,Noginov:2009,Oulton:2009}. Despite the rapid development of numerical simulation techniques, the strength of resonant response of an arbitrary
plasmonic nanostructure is not easy to understand \cite{Bigelow:2012}. In particular, the scattering interaction strength between
electron and photon mediated by plasmon resonance is non-trivial and, meanwhile, very crucial for high resolution microscopy
and spectroscopy on plasmonic nanostructures.

Predicted by an eigen-response theory, dark plasmon modes
\cite{Stockman:2001,Markel:1995,Benisty:2009}) are
considered to be weakly-radiative plasmon modes in nanostructures which can
give high gain factor in stimulated emission \cite{Berini:2012,Noginov:2009}. Research interest in dark plasmon modes and the
associated Fano phenomena \cite{Fano:1961} has been growing rapidly due
to many potential applications such as sensors, lasing, and nonlinear and
slow-light devices \cite{Luk:2010,Giannini:2010,Zhang:2008,Liu:2009,Lassiter:2010,Fan:2010}. Recently, dark plasmon modes have been observed in
optical nanoantennas \cite{Schuck:2005} using electronic excitation
\cite{Abajo:2010,Chu:2009,Koh:2011,Mirsaleh-Kohan:2012}. This opens great
opportunity for using electron beam to study the local strength of high
quality plasmonic resonances in spatial resolution smaller than 10 nm. In
particular, if the dark mode can be observed using cathodoluminescence (CL)
spectroscopy \cite{Yacobi:1990}, it will show great advantages in imaging
plasmonic nanostructures due to its simultaneous high spatial and spectral
resolutions.

Here, we use an eigen-response theory \cite{Markel:1995,Bergman:1980,Fung:2007} to study the strong interaction strength of electron-photon scattering mediated by localized plasmons. The theory predicts a counter-intuitive response from dark plasmon, which leads to a strong scattering interaction between a high
energy electron and a photon. We use finite-difference-time-domain (FDTD)
simulation and CL to verify the theoretical prediction by studying the
scattering between electron and photon close to a plasmonic
nanoantenna. The theoretical predictions agree with our CL experimental results.

We first briefly introduce the prediction from the eigen-response theory.
For a given excitation field ${{\bf E}}_{exc} ({{\bf r}},\omega )$,
the general response polarization (dipole moment density) ${{\bf P}}({{\bf
r}},\omega )$ can be written as a linear combination of the eigenmodes
${{\bf P}}_j ({{\bf r}},\omega )$, where $j$ is a label of one eigenmode.
In an abstract-vector form, it is written as
\begin{equation}
\left| P \right\rangle = \sum\nolimits_j {\alpha _{eig,j} \left|
{P_j } \right\rangle \left\langle {P_j \vert E_{exc} } \right\rangle },
\label{Eq01}\\
\end{equation}

\noindent
where $\left| {P_j } \right\rangle $ and $\lambda _j$ ($\equiv \alpha^{-1}_{eig,j}$) are, respectively, the $j$-th eigenmode
and the $j$-th eigenvalue of an operator $M$ which is defined in the relation between the
excitation field ${{\bf E}}_{exc} ({{\bf r}},\omega )$
and the response ${{\bf P}}({{\bf r}},\omega )$
through $M \left| P \right\rangle = \left| E_{exc} \right\rangle$. Since
$\alpha_{eig,j}$ has a dimension of
polarizability, it is called eigen-polarizability \cite{Fung:2007}
of the $j$-th eigenmode. As we will see below, Eq. [1] suggests that a dark
mode can contribute to a higher detected signal than a bright mode does in
some situations, which seems to be contradictory to our usual believe.

\begin{figure*}[htbp]
\centering
\includegraphics[width=5.4in]{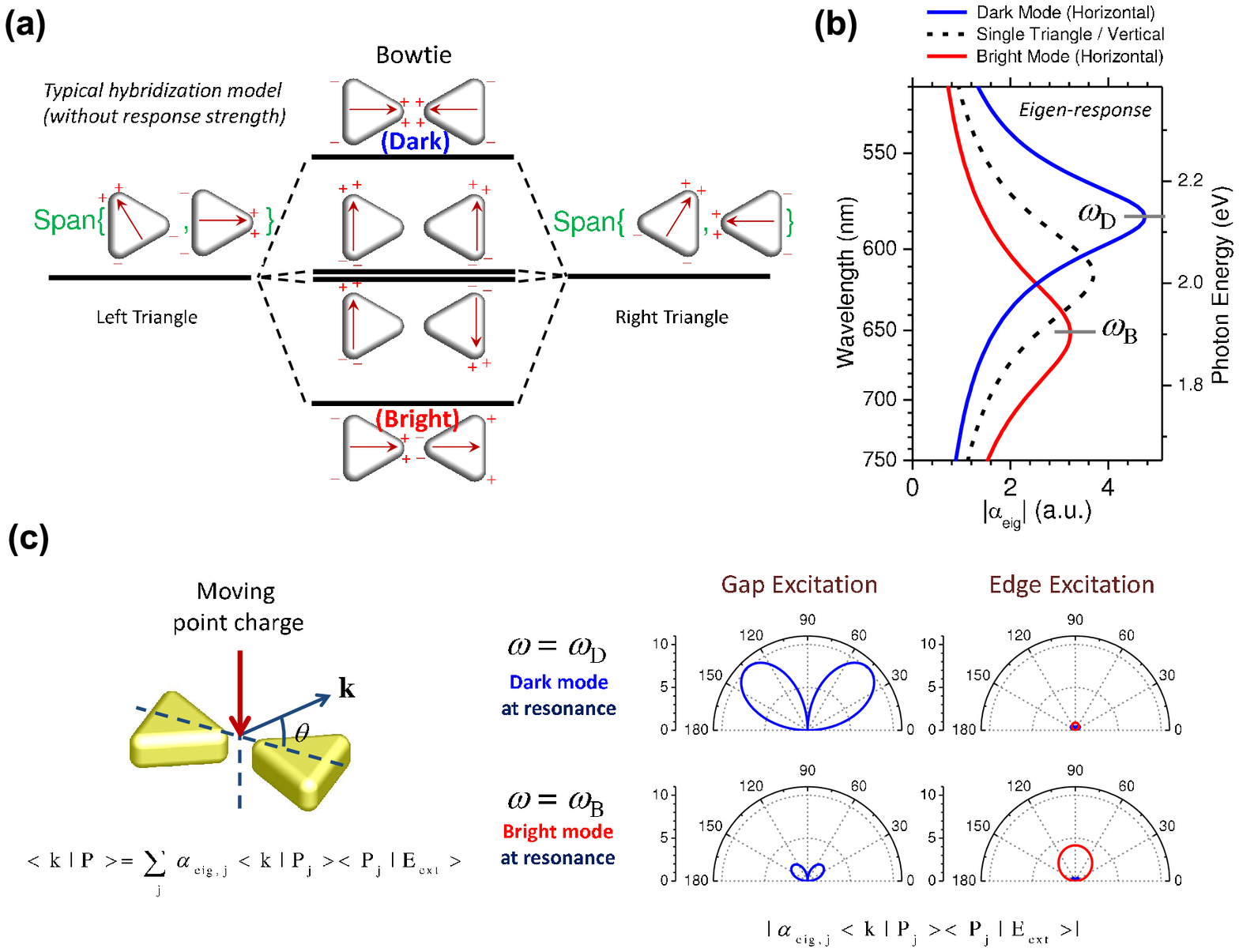}
\caption{\small (color online) Demonstration of Selection rule for CL. (a) Schematic
diagram illustrating the hybridization between left and right particles,
which form the dark modes in bowtie nanoantenna. ``Span'' indicates the
degenerate space spanned by the dipole modes. (b) Magnitudes of the
eigenpolarizabilities, $\alpha_{eig}$, for the hybridized modes. The peaks indicate the resonant
frequencies. (c) Predicted interaction strength between electron beam and
photon as a function of angle $\theta$. The eigen-response theory predicts that the electron-photon scattering interaction strength mediated by dark plasmon can be higher than that of the bright plasmon mode.
Units are arbitrary. Each Au particle has a tip-to-base size of 110 nm and
thickness of 50 nm. Every corner/edge has a rounding radius of 15 nm.}
\end{figure*}

To explain our prediction, let us consider a concrete example (bowtie nanoantenna). Dark modes can be formed in a system of coupled dipole resonators due to the hybridization among dipole modes
\cite{Markel:1995}, such as in bowtie nanoantenna
\cite{Schuck:2005}. The hybridization diagram for a bowtie nanoantenna
formed by two equilateral Au triangles is shown in Figure 1(a) for the dominant
in-plane dipole modes in single triangles. Two-fold degenerate dipole modes
in each triangle hybridize with the modes in the opposite triangle, giving a
total of four hybridized modes which include horizontal dark and bright
modes indicated in Fig. 1(a) and the other almost degenerate vertical modes.
The magnitudes of the eigenpolarizbilities, $\alpha_{eig}$, for these plasmon modes are shown in Fig. 1(b) with
their peaks indicating the resonant frequencies. The horizontal dark and
bright modes are well-separated in frequency while the vertical modes are
almost indistinguishable. In the following, we will focus on distinguishing
the horizontal dark and bright plasmon modes. For a system that supports one
dark mode $\left| P_{D}\right\rangle$ and one
bright mode $\left| P_{B}\right\rangle$, the
radiation amplitude (i.e., the interaction strength with free photon) is
\begin{eqnarray}
&\left\langle k | P \right\rangle
= \alpha_{eig,B}\left\langle k | P_{B} \right\rangle \left\langle P_{B} | E_{exc} \right\rangle \nonumber\\
&+ \alpha_{eig,D}\left\langle k | P_{D} \right\rangle \left\langle P_{D} | E_{exc} \right\rangle,
\label{Eq02}
\end{eqnarray}

In general, $\left\langle k | P_{D} \right\rangle$ has a
magnitude smaller than $\left\langle k | P_{B} \right\rangle$ (see Fig. 1$C)$. However, a dark mode with high quality factor should also have larger magnitude of $\alpha_{eig}$ at
resonance \cite{Fung:2008}). As a result, the magnitude of
$\alpha_{eig,D}\left\langle k | P_{D} \right\rangle$ can be comparable
with $\alpha_{eig,B}\left\langle k | P_{B} \right\rangle$. In addition,
a crucial factor that determines the ultimate radiation is the projection
magnitudes $\left\langle P_{B} | E_{exc} \right\rangle$ and $\left\langle P_{D} | E_{exc} \right\rangle$. By choosing a zero
projection to the bright mode (i.e., making $\left\langle P_{B} | E_{exc} \right\rangle = 0$, we can have strong photon radiation dominated by
the dark mode, which means a strongly enhanced interaction between electron
and photon by dark plasmon. Details of how each of these quantities
contributes to the strong interaction are illustrated in the Supplemental
Material. For an excitation by a high energy electron (30 keV), the final interaction strength predicted by the theory is shown in Fig. 1(c).

To support our prediction, we performed FDTD
simulations as well as experiments for the case of excitation by electron
beam, which is considered to be a fine and controllable excitation source.
In both our simulation and experiment, a 30 keV electron beam is incident
normally to a bowtie nanoantenna. Electron beam is modeled in FDTD
simulation as a moving point charge. Details of simulation method can be
found in Supplemental Material and elsewhere \cite{Chaturvedi:2009}. The
three dimensional geometry of the bowtie antenna in our simulation is almost
the same as in the experiment, except the imperfection of the fabricated
sample and the very thin ($\sim $3nm) adhesion layer below Au particles. To
match the fabricated sample, all corner and edges in the model structure
have a rounding radius of 15 nm.

\begin{figure}[htbp]
\centering
\includegraphics[width=3.2in]{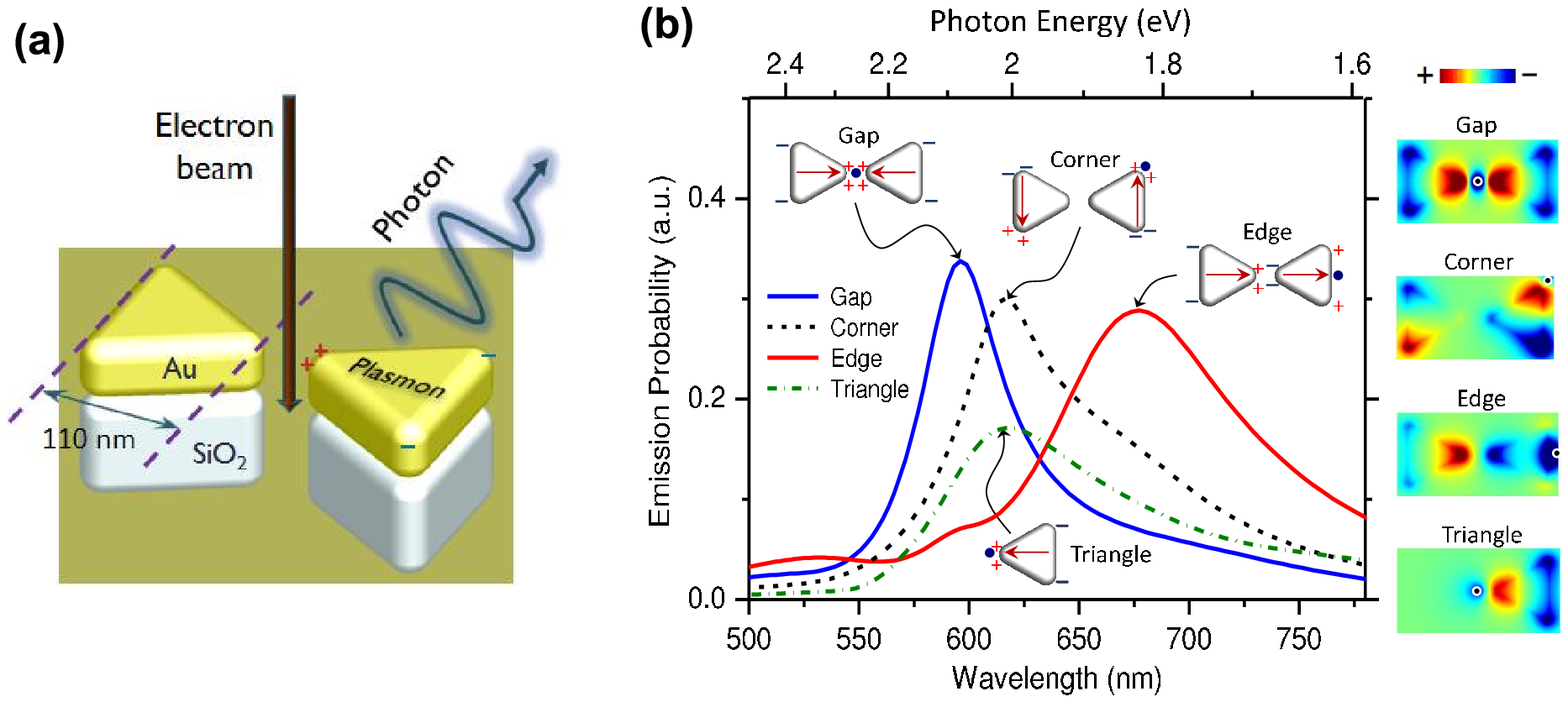}
\caption{\small (color online) Numerical results on the scattering interaction strength between
a 30keV electron and a photon. (a) The rounded bowtie structure used in
numerical simulations. Each Au particle is lifted away from the substrate by
a 85 nm thick $SiO_2$ layer
of the same shape. (b) Solid lines display the photon emission spectra for
bowtie nanoantenna, which indicate the scattering interaction strength. Dashed line displays the spectrum for single triangle.
The diagrams next to spectral peaks indicate the position of electron beams
and the corresponding plasmon modes excited by the beam. Colored plots on
the right show the field patterns on a plane that is 2 nm above the surface
of the nanoantennas. The color indicates the electric field normal to the
monitored plane, which can approximately represent the surface charge
density.}
\end{figure}

Fig. 2(a) shows a schematic of the problem we consider. We simulate and measure
the photon emission for different fixed e-beam locations, which indicates
the strength of the scattering interaction between electron beam and photon.
Simulation results in Fig. 2(b) show that there is a dominant peak associated
with each fixed e-beam. When the e-beam is fixed at the center (gap), right
upper corner, and right edge of the bowtie antenna, we observe a peak at
$\sim $ 600 nm, 620 nm, and 680 nm, respectively. We see that there are at
least three different plasmon modes supported by the bowtie antenna. This is
consistent with our prediction that four resonant modes are supported while
two of the modes are almost degenerate such that they can hardly be
distinguished. To further verify our theory and understand the peaks, we
also simulate the case of single nanotriangle [dashed line in Fig. 2(b)]. Such
a peak wavelength corresponds to the dipole resonance of single nanotriangle
while the peaks for gap, edge, and corner excitations correspond to
anti-parallel horizontal dipoles, parallel horizontal dipoles, and vertical
dipoles, respectively. The above classification of peaks is verified by
simulating the field patterns at the corresponding peaks in Fig. 2(b). The
right colored panels in Fig. 2 show the z-component of the electric field at
a plane located 2 nm above the surface of the bowtie nanoantenna for the
three dominant peaks. These patterns can approximately represent the surface
charge density. We see that, when fixing e-beam at the central gap and
observing the field pattern at the wavelength of 600 nm, the distribution of
the induced charges is symmetric in y-direction, which represents a pair of
anti-parallel dipoles. For edge excitation at 680 nm, the induced charges
show an almost antisymmetric distribution except the field produced by the
e-beam itself near the right edge. For corner excitation at 620 nm, the
induced charges show a pair of anti-parallel vertical dipoles. This agrees
with our theory that the vertical modes have wavelengths very close to the
single triangle case. Apart from the dominant peak positions, we also see
some small features at shorter wavelengths, which may corresponds to higher
order modes.

Here, we briefly discuss why both dark and bright modes can be selectively
excited and analyzed in the far-field with strong signals. When we fix the
e-beam at the center of the gap, the excitation field produced by the e-beam
has an azimuthal symmetry with respect to the center of the bowtie.
Therefore, only the plasmon with charge distribution symmetric in both
directions can be excited ($\left\langle P_{D} | E_{exc} \right\rangle \neq 0$ and $\left\langle P_{B} | E_{exc} \right\rangle = 0$) and
this leads to a pure excitation of horizontal anti-parallel dipole mode,
which has the shortest wavelength among the three observable peaks. When the
e-beam is fixed at the edge, a mirror symmetry is broken and the excitation
of the horizontal parallel dipole mode is possible ($\left\langle P_{D} | E_{exc} \right\rangle \neq 0$ and $\left\langle P_{B} | E_{exc} \right\rangle \neq 0$). Since the e-beam is far away from the center of
bowtie nanoantenna, it is more favorable to the excitation of horizontal
parallel mode, which leads to a peak at the longest wavelength. Similarly,
in the case of e-beam fixed at the corner, the vertical dipole modes can be
excited due to the broken mirror symmetry in the vertical direction. Details
of the projection magnitudes and the emission strengths as a function of the
position of electron beam are described in the Supplemental Material.

The gold bowtie nanoantenna was fabricated using electron-beam lithography
on a multilayered substrate with minimal background luminescence and
relatively low substrate index \cite{Kumar:2010}. In our CL experiment,
an Aluminum parabolic mirror, with a small hole for electron beam, was
placed on top of the sample collected the photons emitted by the antenna
irradiated with an electron beam accelerated at 30 kV and 20 nA current. The
collected photons were directed into a Mach-Czerny type monochromator to
collect spectral information and imaging. Experimental setup have been
previously published with details \cite{Chaturvedi:2009,Kumar:2010}. More information related to this experiment is
included in Supplemental Material. Our experimental results (Fig. 3) show
the strong scattering interaction mediated by the dark plasmon, which agrees
very much with our theory. An SEM picture of the fabricated bowtie
nanoantenna is shown in Fig. 3(a). We observed three peaks for center (gap),
corner, and edge excitations, indicated in the same SEM picture as blue,
black, and red dots, respectively. The observation of the three modes is
consistent with a previous related experiment \cite{Koh:2011}. The
results [Fig. 3(b)] also agree well with the simulation results in terms of the
number of peaks and relative peak positions, except the separation between
peaks are larger in the experiment. The obtained peak wavelengths for
center, corner, and edge excitations are, respectively, 600 nm, 650 nm, and
740 nm. We believe that the discrepancy from simulation results can be due
to the detailed material and geometrical properties. The panchromatic CL
image [Fig. 3(c)] also indicates that the edge excitation gives a weak signal
even the bright mode is excited.

\begin{figure}[htbp]
\centering
\includegraphics[width=3in]{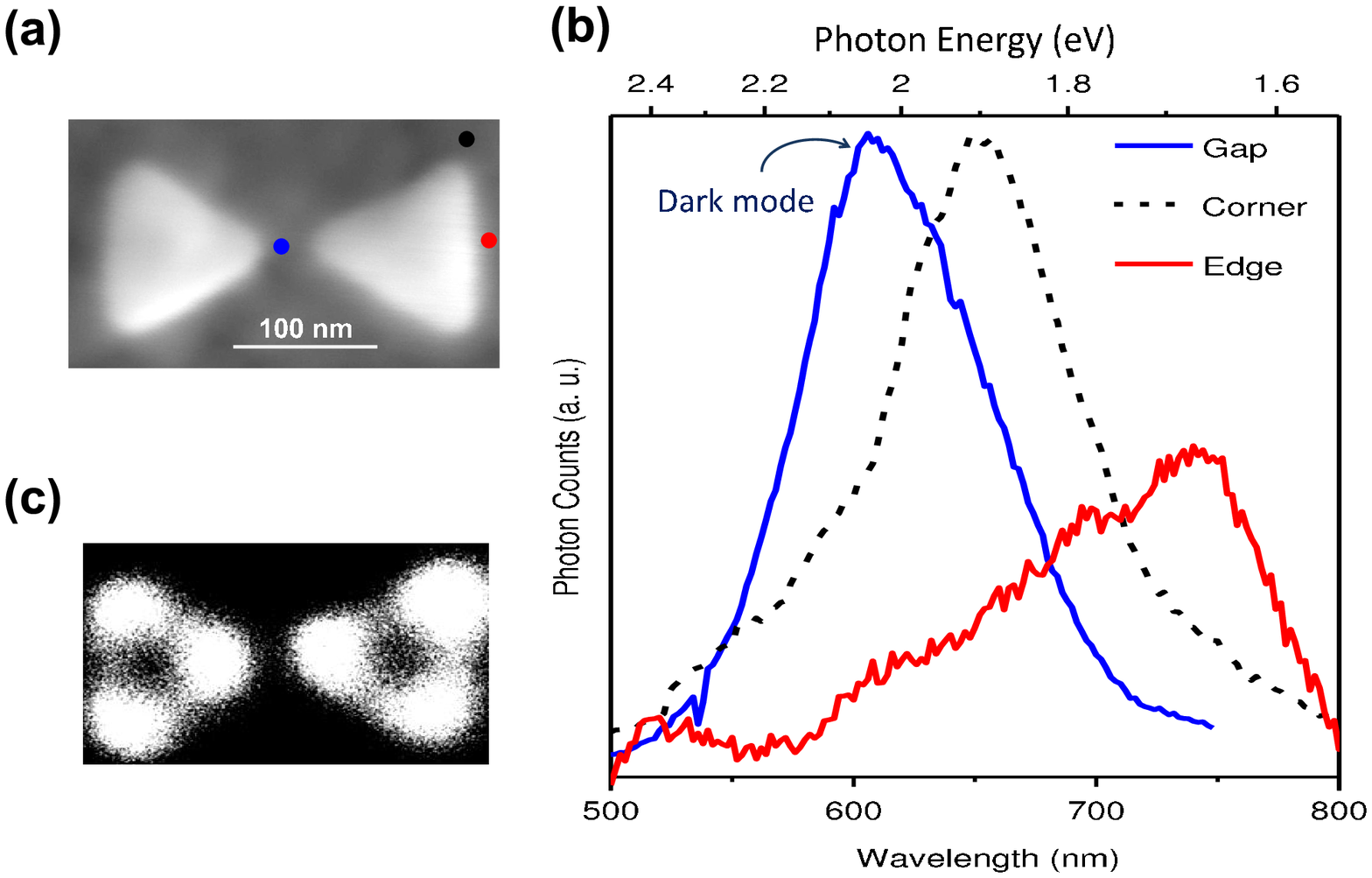}
\caption{\small (color online) Experimentally measured scattering interaction strength at different
selected locations. (a) SEM picture of a fabricated bowtie antenna. (b)
Measured CL spectra at three locations indicated as colored dots in (c).
Panchromatic spatial image collected for the whole spectrum detected by the
photodetector. Bright color corresponds to high photon counts.}
\end{figure}

To further demonstrate the roles of projected magnitudes $\left\langle P_{j} | E_{exc} \right\rangle$, we repeat the simulation by changing
the position of the electron beam from the gap to the edge. The results in
Fig. 4 show the peak positions for different e-beam excitation locations are
the same except the strength of signal, indicating a gradual change in
projection magnitude from domination of anti-parallel mode to parallel mode
(from ``i'' to ``iv''). We see that there is no component of parallel mode
contributing to the response when the e-beam is located at the gap and the
radiation from the anti-parallel mode is thus the only dominant mode
observed. It should be emphasized that the signal for position ``i'' is even
higher than that of the parallel mode for position ``iv'', indicating a
strong interaction between electron beam and photon. This is also observed
in our experimental results in Fig. 3(b).

\begin{figure}[htbp]
\centering
\includegraphics[width=2.4in]{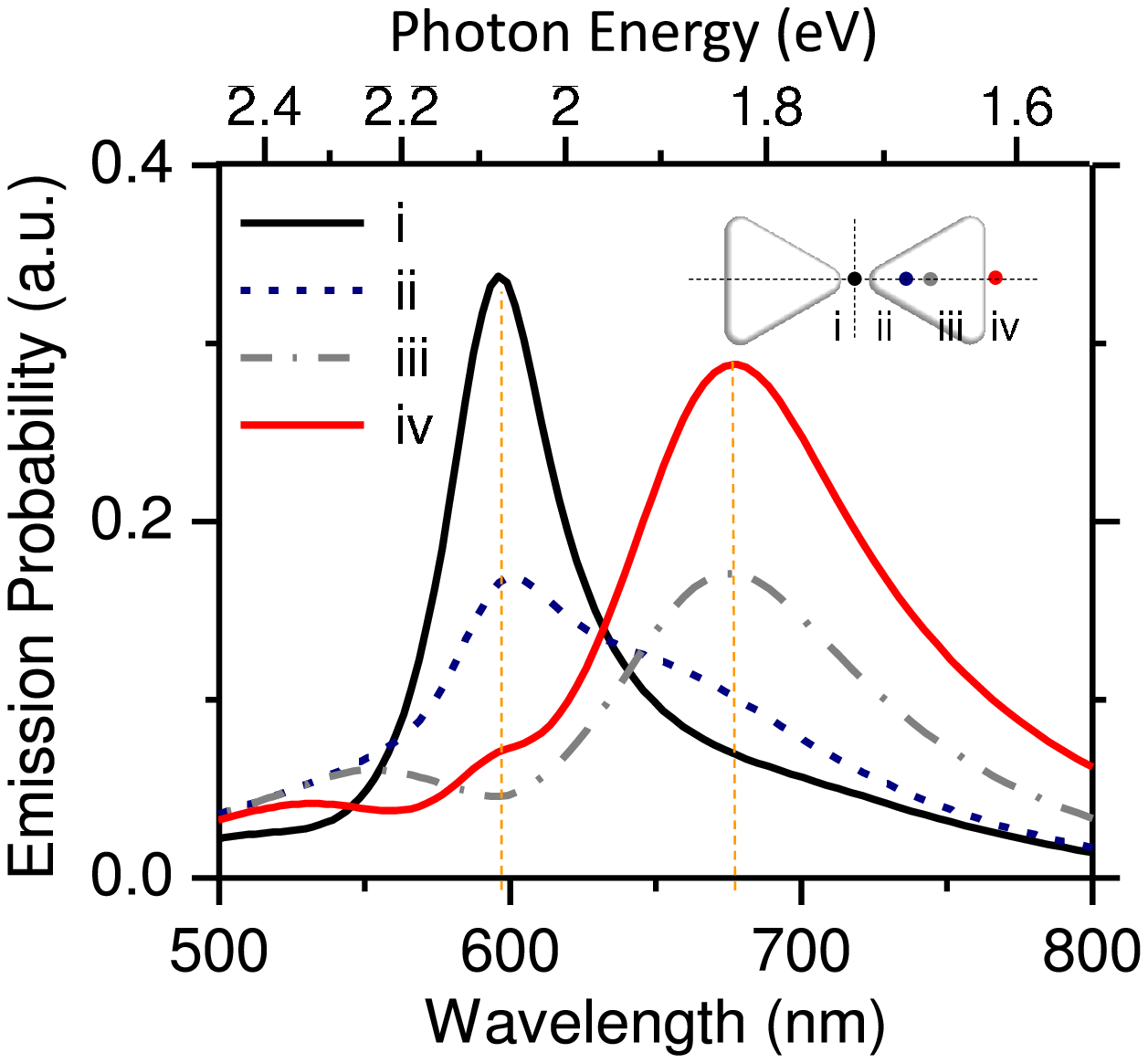}
\caption{\small (color online) Simulated results showing projection magnitudes of the
excitation to bright and dark modes. The locations of electron beam
are shown in the inset. When the electron beam is fixed at i, only the
horizontal anti-parallel dipole modes can be excited (i.e.,
$\left\langle P_{D} | E_{exc} \right\rangle \neq 0$ and
$\left\langle P_{B} | E_{exc} \right\rangle = 0$), which leads to a
single peak close to 600 nm. As we move the electron beam from i to iv, the
horizontal parallel dipole modes ($\sim $ 680 nm) contribute more to the
projection magnitude $\left\langle P_{B} | E_{exc} \right\rangle \neq 0$.}
\end{figure}

To conclude, we introduced a nontrivially strong electron-photon scattering
interaction enhanced by dark plasmon modes. Our theory predicts that even
though dark plasmon mode couples weakly with photon, it can strongly enhance
the scattering interaction between a high energy electron and a photon. Our
simulation and experiment strongly support the theoretical predictions. The
discovery may offer new opportunities for improving single-plasmon generation and
detection in nanostructures. Our study also provides new insights for developing new high-resolution characterization techniques
for high quality plasmonic materials. The phenomenon presented in this
Letter is explained by a classical model. It would be interesting to study
the quantum interaction in the future.

This work was supported by National Science Foundation and
the Office of Naval Research and carried out in part in the Frederick Seitz
Materials Research Laboratory Central Facilities, University of Illinois,
which are partially supported by the U.S. Department of Energy. We thank Dr.
Jun Xu, Dr. Hyungjin Ma, Prof. C. T. Chan, Prof. Lei Zhou, and Prof. Min
Chen for fruitful discussions.

\end{document}